\documentclass[a4paper,12pt]{article}
\usepackage{amssymb}
\usepackage{amsmath}
\usepackage{epsfig}

\usepackage{graphics}

\usepackage{latexsym}
\usepackage{rotating}
\title{ Nonlocal Super Integrable Equations}

\author {
Metin G{\" u}rses \thanks{Email:gurses@fen.bilkent.edu.tr}\\
{\small Department of Mathematics, Faculty of Science} \\
{\small Bilkent University, 06800 Ankara - Turkey}}

\date{\nonumber}
\begin{document}
\maketitle
\date{\nonumber}

\baselineskip 17pt

\numberwithin{equation}{section}

\begin{abstract}
We present nonlocal integrable reductions of super AKNS coupled equations. By the use of nonlocal reductions of Ablowitz and Musslimani we find new super integrable equations. In particular we introduce nonlocal super NLS equations and the nonlocal super mKdV equations.
\end{abstract}

\newtheorem{thm}{Theorem}[section]
\newtheorem{Le}{Lemma}[section]
\newtheorem{defi}{Definition}[section]

\def \part {\partial}
\def \be {\begin{equation}}
\def \ee {\end{equation}}
\def \bea {\begin{eqnarray}}
\def \eea {\end{eqnarray}}
\def \ba {\begin{array}}
\def \ea {\end{array}}
\def \si {\sigma}
\def \al {\alpha}
\def \la {\lambda}
\def \D {\displaystyle}


\section{Introduction}
As nonlocal equations in integrable systems we were accustomed with integro-differential equations. Recently Ablowitz and Musslimani \cite{ab1}-\cite{ab3} have introduced
a new nonlocal reductions of certain integrable coupled differential equations. This approach opened a new channel in nonlocal integrable nonlinear differential equations. The new integrable equations  are either time (T), space (P) or space and time (PT) symmetric. As an example consider AKNS system \cite{abl0},\cite{abl}

\begin{eqnarray}
a\,q_{t}&=& q_{xx}-2 q^2\, p, \label{denk1}\\
a\,p_{t}&=&-p_{xx}+2 p^2\, q. \label{denk2}
\end{eqnarray}
where $p(t,x)$ and $q(t,x)$ are complex dynamical variables and $a$ is a complex number in general. The standard reduction of this is system
is obtained by letting $p(t,x)=k q(t,x)^{\star}$ where $\star$ is the complex conjugation and $k$ is a real constant. With this condition on the dynamical variables $q$ and $p$ the system of equations (\ref{denk1}) and (\ref{denk2}) reduce to the following nonlinear Schrodinger equation (NLS)

\begin{equation}
a q_{t}=a q_{xx}-2k q^2\, q^{\star}, \label{denk3}
\end{equation}
provided that $a^{\star}=-a$. Recently Ablowitz and Musslimani found another integrable reduction of the AKNS system which is given by

\begin{equation}
p(t,x)=k [q(\epsilon_{1} t, \epsilon_{2} x)]^{\star} \label{non}
\end{equation}
where $ (\epsilon_{1})^2=(\epsilon_{2})^2=1$. Under this
condition the above AKNS sytem (\ref{denk1}) and (\ref{denk2}) reduce to

\begin{equation}
a \,q_{t,x}=q_{xx}(t,x) -2 k q^2(t,x)\, [q(\epsilon_{1} t, \epsilon_{2} x)]^{\star}, \label{denk4}
\end{equation}
provided that $a^{\star}=- \epsilon_{1} a$. Nonlocal reductions correspond to $(\epsilon_{1}, \epsilon_{2})=(-1,1),(1,-1),(-1,-1)$. Hence corresponding to these pairs of $\epsilon_{1}$ and  $\epsilon_{2}$ we have three different Nonlocal Integrable NLS equations.
Ablowitz and Musslimani have found also the nonlocal modified KdV equation, nonlocal Davey-Stewartson equation, nonlocal sine-Gordon equation and nonlocal $2+1$ dimensional three-wave interaction equations. Fokas \cite{fok} has introduced nonlocal versions of multidimensional Schrodinger equation, Ma et al., \cite{ma} pointed out that nonlocal complex mKdV equation introduced by Ablowitz and Musslimani is gauge equivalent to a spin-like model. Gerdjikov and Saxena \cite{gerd} studied the complete integrability of nonlocal nonlinear Schrodinger
equation and Sinha and Ghosh \cite{sin} have introduced the nonlocal vector nonlinear Schrodinger equation. Nonlocal system equations were also studied recently  \cite{gerd1}- \cite{gerd3}, \cite{gur2}.

In this work we will consider super AKNS equations \cite{gur0}-\cite{gur1} and give all possible nonlocal reductions of this system. For this purpose in section 2 we start with the super AKNS system where the super potentials are assumed to be polynomials of degree three or less.  In section 3 we study all possible reductions of the super NLS equations. Here we find new super integrable equations, nonlocal super NLS equations. In section 4 we study all possible reductions of the super mKdV equations and find another new super integrable equations, nonlocal super mKdV equations.

\section{super AKNS System}

The lax equations for the super AKNS system are given by \cite{gur0}-\cite{gur1}

\begin{equation}
\Psi_{x}+{\cal A}\, \Psi=0.~~~\Psi_{t}+{\cal B}\, \Psi=0,
\end{equation}
where
\begin{equation}
{\cal A}=\left( \begin{array}{ccc}-i \lambda & r & \beta \\
q & i \lambda & \varepsilon \\
\varepsilon & -\beta & 0
\end{array}\right),~~~
{\cal B}= \left(\begin{array}{ccc} A & C & \alpha \\
B & -A & \rho \\
\rho & -\alpha &0
\end{array} \right)
\end{equation}
where the super functions  $A, B, C$ are commuting and $\alpha, \rho$ are anti-commuting fields. The super AKNS dynamical variables are
$q,r$ and $\beta, \varepsilon$. Here $q$ and $r$ are bosonic and $\beta$ and $\varepsilon$ are the fermionic variables. The super functions $A,B,C$ and $\alpha, \rho$ are functions of $t$ and $x$ through the dynamical variables and their derivatives with respect to $x$. The super functions depend also on the spectral variable $\lambda$ but the dynamical variables do not depend on it. Considering that the super functions are polynomial functions of $\lambda$ of degree three and using the integrability condition $\Psi_{tx}=\Psi_{xt}$ we obtain the following super integrable equations.

\begin{eqnarray}
&& q_{t}=a_{2}\, (-\frac{1}{2}\, q_{xx}+q ^2\, r+2 \, \varepsilon_{x}\, \varepsilon+2 q \beta \varepsilon)+i a_{3}\,(-\frac{1}{4}\, q_{xxx}+\frac{3}{2}\, q r q_{x}+3 (\varepsilon_{x} \varepsilon)_{x} \nonumber \\
&&-3 q \beta_{x} \varepsilon +3 q \beta \varepsilon_{x}), \label{sup1} \\
&& r_{t}=a_{2}\, (\frac{1}{2}\, r_{xx}-q \, r^2 +2\, \beta_{x}\, \beta-2 r \beta \varepsilon)+i a_{3}\,(-\frac{1}{4}\, r_{xxx}+\frac{3}{2}\, q r r_{x}-3 (\beta_{x} \beta)_{x} \nonumber \\
&&+3 r \beta_{x} \varepsilon -3 r \beta \varepsilon_{x}), \label{sup2} \\
&&\beta_{t}=a_{2}\,(\beta_{xx}-r \varepsilon_{x}-\frac{1}{2}\, \varepsilon r_{x}-\frac{1}{2} q r \beta)+i a_{3} (-\beta_{xxx}+\frac{3}{4} r q_{x} \beta+\frac{3}{4} q r_{x} \beta+\frac{3}{2} q r \beta_{x} \nonumber \\
&&+\frac{3}{2} r_{x} \varepsilon_{x}+\frac{3}{4} \varepsilon r_{xx}) \label{sup3}\\
&&\varepsilon_{t}=a_{2}\,(-\varepsilon_{xx}+q \beta_{x}+\frac{1}{2}\, \beta q_{x}+\frac{1}{2} q r \varepsilon)+i a_{3} (-\varepsilon_{xxx}+\frac{3}{4} r q_{x} \varepsilon+\frac{3}{4} q r_{x} \varepsilon +\frac{3}{2} q r \varepsilon_{x} \nonumber \\
&&+\frac{3}{2} q_{x} \beta_{x}+\frac{3}{4} \beta q_{xx}) \label{sup4}
\end{eqnarray}
where $a_{2}$ and $a_{3}$ are arbitrary constants. If $a_{3}=0, a_{2} \ne 0$ the above system is the coupled super nonlinear Schrodinger equations (sNLSE) and if $a_{2}=0, a_{3} \ne 0$ the above system is the coupled super modified Korteweg de Vries equations (smKdV). Next sections we investigate the standard and nonlocal reductions of these equations.

\section{Nonlocal super NLS system}

In this and next section we investigate  the possibility of nonlocal reductions in the system of super AKNS equations. For this purpose we start with simplest case, super NLS equations.
Letting $a_{3}=0$ in the above equations (\ref{sup1})-(\ref{sup4}) we get the super coupled NLS equations. There are two bosonic ($p$,$r$) and two fermionic ($\varepsilon, \beta$) potentials satisfying

\begin{eqnarray}
a q_{t}&=&-\frac{1}{2}\,q_{xx}+q^2\,r+2 \varepsilon_{x}\, \varepsilon+2 q\, \beta\, \varepsilon, \label{seq1}\\
a r_{t}&=&\frac{1}{2}\,r_{xx}-q\,r^2+2 \beta_{x}\, \beta-2 r\, \beta\, \varepsilon, \label{seq2}\\
a \varepsilon_{t}&=&-\varepsilon_{xx}+q\, \beta_{x}+\frac{1}{2}\, \beta \, q_{x}+\frac{1}{2}\,q\,r \,\varepsilon, \label{seq3}\\
a \beta_{t}&=&\beta_{xx}-r\, \varepsilon_{x}-\frac{1}{2}\, \varepsilon \, r_{x}-\frac{1}{2}\,q\,r \,\beta, \label{seq4}
\end{eqnarray}
where $a_{2}=1/a$.
The standard reduction is $r=k_{1}\, \bar{q}$ and $\beta=k_{2} \bar{\varepsilon}$ where  $k_{1}$ and $k_{2}$ are constant, a bar over a
quantity denotes the Berezin conjugation in the Grassman algebra. If ${\cal P}$ and ${\cal Q}$ are super functions (bosonic or fermionic) then
$\overline{{\cal P}{\cal Q}}={\overline{\cal Q}}\,{ \overline{\cal P}}$. Under these constraints the above equations (\ref{seq1})-(\ref{seq4}) reduce to the following super NLS equations \cite{kup},\cite{kup1} provided $k_{1}=k_{2}^2$ and $\bar{a}=-a$

\begin{eqnarray}
a q_{t}&=&-\frac{1}{2}\,q_{xx}+k_{1}\, q^2\,\bar{q}+2 \varepsilon_{x}\, \varepsilon+2 k_{2}\,q\, \bar{\varepsilon}\, \varepsilon, \label{seq5}\\
a \varepsilon_{t}&=&-\varepsilon_{xx}+k_{2}\, q\, \bar{\varepsilon}_{x}+\frac{1}{2}\, k_{2}\, \bar{\varepsilon} \, q_{x}+ \frac{1}{2}\,k_{1}\,q\, \bar{q} \,\varepsilon, \label{seq6}
\end{eqnarray}
For the usual complex valued functions, Berezin conjugations reduces to a complex conjugation.

In this work we show that super NLS system (\ref{seq1})-(\ref{seq4}) can be reduced to nonlocal super NLS equations. This can be done by
 choosing the super Ablowitz-Musslimani reduction as

 \begin{equation}
 r(t,x)=\epsilon\, \bar{q}(\epsilon_{1} t, \epsilon_{2} x),~~~ \beta(t,x)= \mu \bar{\varepsilon}(\epsilon_{1} t, \epsilon_{2} x).
 \end{equation}
 where $\epsilon_{1}^2=\epsilon_{2}^2=1$.
Under these constraints the above set (\ref{seq1})-(\ref{seq4}) reduces to

 \begin{eqnarray}
a q_{t}(t,x)&=&-\frac{1}{2}\,q_{xx}(t,x)+\epsilon\,\, q^2(t,x)\,\bar{q}(\epsilon_{1} t, \epsilon_{2} x)+2 \varepsilon_{x}(t,x)\, \varepsilon(t,x) \\
&&+2 \mu\,q\, \bar{\varepsilon}(\epsilon_{1} t, \epsilon_{2} x), \varepsilon(t,x), \label{seq7}\\
a \varepsilon_{t}(t,x)&=&-\varepsilon_{xx}(t,x)+\mu\, q(t,x)\, \bar{\varepsilon}_{x}(\epsilon_{1} t, \epsilon_{2} x)+\frac{1}{2}\, \mu \, \bar{\varepsilon}(\epsilon_{1} t, \epsilon_{2} x) \, q_{x}(t,x) \\
&&+ \frac{1}{2}\,\epsilon\,q(t,x)\, \bar{q}(\epsilon_{1} t, \epsilon_{2} x) \,\varepsilon(t,x),  \label{seq8}
\end{eqnarray}
provided that
\begin{equation}
\bar{a}\, \epsilon_{1}=-a,~~~ \mu^2\, \epsilon_{2}=\epsilon.
\end{equation}
Nonlocal reductions correspond to the choices $(\epsilon_{1},\epsilon_{2})=(-1,1),(1,-1),(-1,-1)$. They are explicitly given by

\vspace{0.3cm}
\noindent
1. T-symmetric nonlocal super NLS equation:

\begin{eqnarray}
a q_{t}(t,x)&=&-\frac{1}{2}\,q_{xx}(t,x)+\epsilon\,\, q^2(t,x)\,\bar{q}(- t, x)+2 \varepsilon_{x}(t,x)\, \varepsilon(t,x) \\
&&+2 \mu\,q\, \bar{\varepsilon}(-t,x), \varepsilon(t,x), \label{seq13}\\
a \varepsilon_{t}(t,x)&=&-\varepsilon_{xx}(t,x)+\mu\, q(t,x)\, \bar{\varepsilon}_{x}(-t,x)+\frac{1}{2}\, \mu \, \bar{\varepsilon}(-t,x) \, q_{x}(t,x) \\
&&+ \frac{1}{2}\,\epsilon\,q(t,x)\, \bar{q}(-t,x) \,\varepsilon(t,x),  \label{seq14}
\end{eqnarray}
with $a^{\star}=a$ and $\epsilon=\mu^2$.

\vspace{0.3cm}
\noindent
2. P-symmetric nonlocal super NLS equation:

\begin{eqnarray}
a q_{t}(t,x)&=&-\frac{1}{2}\,q_{xx}(t,x)+\epsilon\,\, q^2(t,x)\,\bar{q}(t, -x)+2 \varepsilon_{x}(t,x)\, \varepsilon(t,x) \\
&&+2 \mu\,q\, \bar{\varepsilon}(t,-x), \varepsilon(t,x), \label{seq15}\\
a \varepsilon_{t}(t,x)&=&-\varepsilon_{xx}(t,x)+\mu\, q(t,x)\, \bar{\varepsilon}_{x}(t,-x)+\frac{1}{2}\, \mu \, \bar{\varepsilon}(t,-x) \, q_{x}(t,x) \\
&&+ \frac{1}{2}\,\epsilon\,q(t,x)\, \bar{q}(t,-x) \,\varepsilon(t,x),  \label{seq16}
\end{eqnarray}
with $a^{\star}=-a$ and $\epsilon=-\mu^2$.

\vspace{0.3cm}
\noindent
3. PT-symmetric nonlocal super NLS equation:

\begin{eqnarray}
a q_{t}(t,x)&=&-\frac{1}{2}\,q_{xx}(t,x)+\epsilon\,\, q^2(t,x)\,\bar{q}(-t, -x)+2 \varepsilon_{x}(t,x)\, \varepsilon(t,x) \\
&&+2 \mu\,q\, \bar{\varepsilon}(-t,-x), \varepsilon(t,x), \label{seq15}\\
a \varepsilon_{t}(t,x)&=&-\varepsilon_{xx}(t,x)+\mu\, q(t,x)\, \bar{\varepsilon}_{x}(-t,-x)+\frac{1}{2}\, \mu \, \bar{\varepsilon}(-t,-x) \, q_{x}(t,x) \\
&&+ \frac{1}{2}\,\epsilon \,q(t,x)\, \bar{q}(t,-x) \,\varepsilon(t,x),  \label{seq16}
\end{eqnarray}
with $a^{\star}=a$ and $\epsilon=-\mu^2$.

\vspace{0.3cm}
\noindent

\section{Nonlocal super mKdV system}

Another special case of the super AKNS equations is the super mKdV equations \cite{gur0}, \cite{gur}.

\begin{eqnarray}
a q_{t}&=&-\frac{1}{4}\,q_{xxx}+\frac{3}{2}\,r\,q\,q_{x}+3 (\varepsilon_{x}\, \varepsilon)_{x}-3\, q\, \beta_{x}\, \varepsilon+3 q \beta\, \varepsilon_{x}, \label{seq9}\\
a r_{t}&=&-\frac{1}{4}\,r_{xxx}+\frac{3}{2}\,r\,q\,r_{x}-3 (\beta_{x}\, \beta)_{x}+3\, r\, \beta_{x}\, \varepsilon-3 r \beta\, \varepsilon_{x}, \label{seq10},\\
a \varepsilon_{t}&=&-\varepsilon_{xxx}+ \frac{3}{4}\, (r\, q)_{x}\, \varepsilon+\frac{3}{2}\,q\,r\, \varepsilon_{x}+\frac{3}{2}\,q_{x}\, \,\beta_{x}+\frac{3}{4} \beta\, q_{xx}, \label{seq11}\\
a \beta_{t}&=&-\beta_{xxx}+ \frac{3}{4} (r\, q)_{x}\, \beta+\frac{3}{2}\,q\,r\, \beta_{x}+\frac{3}{2}\,r_{x}\, \,\varepsilon_{x}+\frac{3}{4} \varepsilon\, r_{xx}, \label{seq12}
\end{eqnarray}
The standard reduction is $r=k_{1} \bar{q}, \beta=k_{2}\, \bar{\varepsilon}$, Then we obtain (\cite{gur0}, \cite{gur})

\begin{eqnarray}
a q_{t}&=&-\frac{1}{4}\,q_{xxx}+\frac{3}{2}\,k_{1}\, \bar{q}\,q\,q_{x}+3 (\varepsilon_{x}\, \varepsilon)_{x}-3\,k_{2} q\, \bar{\varepsilon}_{x}\, \varepsilon+3 k_{2}\,q \bar{\varepsilon}\, \varepsilon_{x}, \label{seq13}\\
a \varepsilon_{t}&=&-\varepsilon_{xxx}+ \frac{3}{4}\,k_{1} (\bar{q}\, q)_{x}\, \varepsilon+\frac{3}{2}\,k_{1}q\, \bar{q}\, \varepsilon_{x}+\frac{3}{2}\,k_{2}\, q_{x}\, \,\bar{\varepsilon}_{x}+\frac{3}{4}\,k_{2} \bar{\varepsilon}\, q_{xx}, \label{seq14}
\end{eqnarray}
provided that $k_{1}=k_{2}^2$ and $\bar{a}=a$.

For the super mKdV system Ablowiz-Musslimani type of reduction is also possible. Letting

\begin{equation}
 r(t,x)=\epsilon\, \bar{q}(\epsilon_{1} t, \epsilon_{2} x),~~~ \beta(t,x)= \mu \bar{\varepsilon}(\epsilon_{1} t, \epsilon_{2} x).
 \end{equation}
 where $\epsilon_{1}^2=\epsilon_{2}^2=1$. We get the following system of equations

 \begin{eqnarray}
a q_{t}(t,x)&=&-\frac{1}{4}\,q_{xxx}(t,x)+\frac{3}{2}\,\epsilon\, \bar{q}(\epsilon_{1} t, \epsilon_{2} x)\,q(t,x)\,q_{x}(t,x)
+3 (\varepsilon_{x}(t,x)\, \varepsilon(t,x))_{x} \nonumber \\
&&-3\, q\, \bar{\varepsilon}_{x}(\epsilon_{1} t, \epsilon_{2} x)\, \varepsilon(t,x)+3 \mu\,q(t,x)\, \bar{\varepsilon}(\epsilon_{1} t, \epsilon_{2} x)\, \varepsilon_{x}(t,x), \label{seq17}\\
a \varepsilon_{t}(t,x)&=&-\varepsilon_{xxx}(t,x)+ \frac{3}{4}\,\epsilon (\bar{q}(\epsilon_{1} t, \epsilon_{2} x)\, q(t,x))_{x}\, \varepsilon(t,x)+\frac{3}{2}\,\epsilon q(t,x)\, \bar{q}(\epsilon_{1} t, \epsilon_{2} x)\, \varepsilon_{x}(t,x)+ \nonumber\\
&&\frac{3}{2}\,\mu\, q_{x}(t,x)\, \,\bar{\varepsilon}_{x}(\epsilon_{1} t, \epsilon_{2} x)
+\frac{3}{4}\,\mu \bar{\varepsilon}(\epsilon_{1} t, \epsilon_{2} x)\, q_{xx}(t,x), \label{seq18}
\end{eqnarray}
provided that $\bar{a}\, \epsilon_{1}\, \epsilon_{2}=a$, $\mu^2 \epsilon_{2}=\epsilon$.
Nonlocal reductions correspond to the choices $(\epsilon_{1},\epsilon_{2})=(-1,1),(1,-1),(-1,-1)$. They are explicitly given by

\vspace{0.3cm}
\noindent
1. T-symmetric nonlocal super NLS equation: $\bar{a}=-a$ and $\epsilon=\mu^2$.

\begin{eqnarray}
a q_{t}(t,x)&=&-\frac{1}{4}\,q_{xxx}(t,x)+\frac{3}{2}\,\epsilon\, \bar{q}(-t,x)\,q(t,x)\,q_{x}(t,x)
+3 (\varepsilon_{x}(t,x)\, \varepsilon(t,x))_{x} \nonumber \\
&&-3\, q\, \bar{\varepsilon}_{x}(-t,x)\, \varepsilon(t,x)+3 \mu\,q(t,x) \bar{\varepsilon}(-t,x)\, \varepsilon_{x}(t,x), \label{seq119}\\
a \varepsilon_{t}(t,x)&=&-\varepsilon_{xxx}(t,x)+ \frac{3}{4}\,\epsilon (\bar{q}(-t,x)\, q(t,x))_{x}\, \varepsilon(t,x)+\frac{3}{2}\,\epsilon q(t,x)\, \bar{q}(-t,x)\, \varepsilon_{x}(t,x)+ \nonumber\\
&&\frac{3}{2}\,\mu\, q_{x}(t,x)\, \,\bar{\varepsilon}_{x}(-t,x)
+\frac{3}{4}\,\mu \bar{\varepsilon}(-t,x)\, q_{xx}(t,x), \label{seq20}
\end{eqnarray}

\vspace{0.3cm}
\noindent
2. P-symmetric nonlocal super NLS equation: $\bar{a}=-a$ and $\epsilon=-\mu^2$.

\begin{eqnarray}
a q_{t}(t,x)&=&-\frac{1}{4}\,q_{xxx}(t,x)+\frac{3}{2}\,\epsilon\, \bar{q}(t,-x)\,q(t,x)\,q_{x}(t,x)
+3 (\varepsilon_{x}(t,x)\, \varepsilon(t,x))_{x} \nonumber \\
&&-3\, q\, \bar{\varepsilon}_{x}(t,-x)\, \varepsilon(t,x)+3 \mu\,q(t,x) \bar{\varepsilon}(t,-x)\, \varepsilon_{x}(t,x), \label{seq119}\\
a \varepsilon_{t}(t,x)&=&-\varepsilon_{xxx}(t,x)+ \frac{3}{4}\,\epsilon (\bar{q}(t,-x)\, q(t,x))_{x}\, \varepsilon(t,x)+\frac{3}{2}\,\epsilon q(t,x)\, \bar{q}(t,-x)\, \varepsilon_{x}(t,x)+ \nonumber\\
&&\frac{3}{2}\,\mu\, q_{x}(t,x)\, \,\bar{\varepsilon}_{x}(t,-x)
+\frac{3}{4}\,\mu \bar{\varepsilon}(t,-x)\, q_{xx}(t,x), \label{seq20}
\end{eqnarray}

\vspace{0.3cm}
\noindent
3. PT-symmetric nonlocal super NLS equation: $\bar{a}=a$ and $\epsilon=-\mu^2$.

\begin{eqnarray}
a q_{t}(t,x)&=&-\frac{1}{4}\,q_{xxx}(t,x)+\frac{3}{2}\,\epsilon\, \bar{q}(-t,-x)\,q(t,x)\,q_{x}(t,x)
+3 (\varepsilon_{x}(t,x)\, \varepsilon(t,x))_{x} \nonumber\\
&&-3\, q\, \bar{\varepsilon}_{x}(-t,-x)\, \varepsilon(t,x)+3 \mu\,q(t,x) \bar{\varepsilon}(-t,-x)\, \varepsilon_{x}(t,x), \label{seq119}\\
a \varepsilon_{t}(t,x)&=&-\varepsilon_{xxx}(t,x)+ \frac{3}{4}\,\epsilon (\bar{q}(-t,-x)\, q(t,x))_{x}\, \varepsilon(t,x)+\frac{3}{2}\,\epsilon q(t,x)\, \bar{q}(-t,-x)\, \varepsilon_{x}(t,x)+ \nonumber\\
&&\frac{3}{2}\,\mu\, q_{x}(t,x)\, \,\bar{\varepsilon}_{x}(-t,-x)
+\frac{3}{4}\,\mu \bar{\varepsilon}(-t,-x)\, q_{xx}(t,x), \label{seq20}
\end{eqnarray}

\section{Conclusion}

In this work we extended the recently found nonlocal reductions of integrable nonlinear equations  to super integrable equations. Starting with the
super AKNS system we studied all possible nonlocal reductions  and found two new integrable systems. They are nonlocal super NLS equations and super mKdV systems of equations. There are three different nonlocal types of super integrable equations. They correspond to $T$-, $P$-, and $PT$- symmetric super NLSE and super mKdV equations. The nonlocal reductions presented in this paper can be applied to any super integrable systems and obtain new super integrable equations.

\vspace{1cm}

\section{Acknowledgment}
  This work is partially supported by the Scientific
and Technological Research Council of Turkey (T\"{U}B\.{I}TAK).

\end{document}